# On the Probability distribution function of particle density at the edge of fusion devices

F. Sattin[#], N. Vianello, M. Valisa

*Consorzio RFX, Associazione Euratom-ENEA per la Fusione,*

*Corso Stati Uniti 4, 35127 Padova, ITALY*

**Abstract**

The Probability Distribution Function (PDF) $P_n(n)$ of the particle density at the edge of several magnetic fusion devices, including tokamaks, stellarators, and linear devices, is known to be strongly non-gaussian. In this paper we present experimental results from RFX Reversed Field Pinch [G. Rostagni, Fus. Eng. Design **25**, 301 (1995)], confirming the universal shape of $P_n$ also for RFP's. An explanation for the form of $P_n$ is attempted, on the basis of simple conservation equations. The model result is shown to fitted fairly well empirical data in a few different experimental scenarios.

---

[#] Corresponding author. E-mail address: fabio.sattin@igi.cnr.it

# I. Introduction

Transport in magnetically confined plasmas is known to be strongly influenced by turbulent processes at the edge. Hence, investigating the nature and the origin of the turbulence is mandatory for performance improvements in these devices.

It is known since years that edge turbulence is mainly electrostatic. A statistical analysis of main edge plasma parameters (density, potential, temperature, …) reveals that they exhibit large bursty fluctuations, dubbed "intermittency". Attention rapidly moved to the study of these bursty events, owing to their importance for the transport: an investigation in the Reversed Field Pinch (RFP) configuration showed that, although relatively rare, they carry more than 50% of total flux losses[1]. Similar figures are valid also for tokamaks[2,3] and linear devices[4]. These intermittent events are nowadays associated to coherent structures moving into a background plasma.

Indeed, although bidimensional arrays of probes are being increasingly popular in order to get snapshots of the full spatial structure of these structures[5,6], most of the investigations are still being carried on through high-sampling-frequency point-wise measurements using Langmuir probes. The data are collected in the form of a time series, and then analyzed using various statistical tools.

Because of its direct connection with transport, particle flux $\Gamma$ attracts a large fraction of investigations. Under the hypothesis of electrostatic turbulence, $\Gamma \propto n \nabla \phi$, with $n$ density and $\phi$ plasma potential. Hence, it is these two quantities that are ultimately being studied. Particle density PDFs ($P_n$), in particular, were studied by Antar $et\ al$ in a number of devices, showing their universal shape, valid for all experiments[4,7,8]. Similar results were found also by Budaev $et\ al$[9].

Usually, one is not interested as much to the absolute value of one parameter $X$ as to its fluctuation: empirically, to the quantity $\delta X = X - <X>$, where $<...>$ stands for the mean value over the experimental points. The parameter $X$ here may stand for plasma potential, temperature, magnetic field, flux, or particle density. In the lack of underlying physics mechanisms which correlate fluctuations between them, different measurements of $X$ are independent between them and hence expected to yield a gaussian PDF for $\delta X$. Experimentally, however, things are different: by example, the issue of the departure from normality was extensively investigated for potential and magnetic field fluctuations within the long-standing debate about the ultimate origin of edge turbulence (Self Organized Criticality-SOC[10-13]-or fluid turbulence[14,15]).

In this work we will focus on density, whose fluctuations are strongly non-gaussian, too. This, together with the already mentioned universal features between different experiments, calls for the existence of some underlying mechanism of general validity, whose investigation could be of importance.

The purpose of this work is threefold:

a) first of all, we will present density data from the Reversed Field Experiment (RFX)[16]. Their PDF very accurately reproduces the shape found from Antar *et al*, thus adding another confirmation to its "universal" behaviour (i.e., independent of the magnetic configuration). We will mainly refer to data collected by using "standard" Langmuir probes, but will also make a brief reference to independent measurements taken using filters monitoring the CII 5150 Å line. Although, as explained later, these data are not suitable candidates to our analysis, they do seem to qualitatively corroborate our results. Spectroscopic diagnostics are not routinely used to monitor turbulence, hence the data presented here are an interesting example of the potential capabilities of this kind of measurements.

b) Second, some comments are advanced about the requirements that any physically meaningful $P_n$ must fulfil. In the light of these remarks, the suggested $P_n$ by Antar *et al* is re-examined; an alternative PDF is suggested and compared against real data.

c) Finally, an interpretation for this PDF is attempted in terms of basic physics.

## II. Measurements

*A. Langmuir probes*

The experimental set-up used for these measurements has been described in detail elsewhere: see, e.g., Ref. [17]. Here, we limit to some basic informations: the sampling frequency of the probe was 1 MHz, with a total of about 40,000 measurement points for each shot. It is well known that density measurement using Langmuir probes can be potentially spoiled by the fact that one is not actually measuring density but the saturation current: $I_s \propto n \times T^{1/2}$. However, in the light of the square root dependence and of the smallness of temperature fluctuations in RFX, we are allowed to neglect the temperature contribution. Data were collected into the RFX Scrape Off Layer (SOL).

The output time series was binned to give the PDF of Fig. 1. This figure can be compared against Fig. 4 of Ref. [7] or Fig. 3 of Ref. [8]. The agreement is remarkable.

We want to stress that, in the Figure, we have plotted the PDF of the *density n*, not of its *fluctuations δn*. The difference may appear trivial, summing up numerically to an offset. However, conceptually, it makes a difference: indeed, along this paper, we will attempt to show that it is the former, not the latter, the physically relevant variable.

*B. Spectroscopic filters*

Filters centered around the 5150 Å wavelength monitored the corresponding CII line evolution at four spatial locations. Passive spectroscopy can only give line-of-sight integrated measurements, but owing to typical RFX temperature and density profiles, CII is likely to exist only at the very edge. Sampling frequency is here just $f_{filters}$ = 1/8 MHz. This, together with the issue of stationarity of the data, limited the amount of samples available to rather short time series: typically, 2-3000 points for each shot.

Under the hypothesis of detailed balance, the number of emitted 5150 Å photons is equal to the number of electron collisional excitations from the ground state: $I$(5150 Å) = $n_{CII} \times n_e \times Q(T)$. Here, again, we have a mixed dependence from plasma density $n_e$ as well as from the temperature $T$ through the rate coefficient $Q$; it has to be simplified under the supplementary hypotheses of small $T$ fluctuations and/or weak $T$ dependence in $Q$ (both hypotheses rather well fulfilled at the edge of RFX). Hence, we choose to omit once again the $T$ dependence and identify fairly simply $I$(5150 Å) $\leftrightarrow n_e$. Of course, impurity density fluctuations must be discarded too in order to reach this result, but-due to the huge inertia of impurities with respect to electrons-this does not appear a stringent bound.

Because of the reduced statistics in comparison with Langmuir probes, only the region around the maximum of the PDF can be accurately sampled. It is displayed in Fig. 2. Even in this case we are plotting the full signal, not its deviation from the mean.

Since the sampling frequency is just slightly above one hundred kHz, and is therefore partially overlapping the range of frequencies containing appreciable amounts of turbulent signal, there is the question to what extent the resulting PDF is perturbed by this effective low-pass filtering procedure. We attempted to have an insight by the following procedure: we filtered the Langmuir signal of Fig. 1, discarding the frequency content above $f_{\text{filters}}$, and recomputed its PDF. The result is that is the high-signal region to be mostly affected: the PDF goes to zero much more steeply than the unfiltered signal, and its slope is very well approximated by an exponential. The top of the curve is instead not dramatically varying. This is consistent with a picture in which the bulk of the signal is given by slow events, and only the tails of the PDF are determined by rapid fluctuations. These consideration can be only qualitative, but give us confidence that the PDF in Fig. 2 maintains some features of the true PDF. It clear, however, that any really accurate analysis cannot rely on the data in Fig. 2: besides the already mentioned issues about statistics, there is the obvious fact that spectroscopy measures *electron* density fluctuations, while Langmuir probes do measure *ion* density fluctuations. Although the two are related, any attempt of comparing results from the two diagnostics is not straightforward.

### III. Some considerations on the analytical form for $P_n$

Let us dwell about the issue of the analytical form for $P_n$ and the related issue of the differences in considering $n$ rather than $\delta n$.

No true investigations have appeared about the analytical functional form for $P_n$: to the best of our knowledge, the only approximation for it is that given by Antar *et al*[8]: a gaussian in the region of negative $\delta n$ and a decaying exponential in the positive region: $P_n(\delta n) = f(\delta n)$, with

$$f(x) = \begin{cases} C \exp\left(-\frac{1}{2}\left(\frac{x-x0}{\sigma}\right)^2\right) & x < x_0 \\ C \exp\left(-\frac{x-x0}{\lambda}\right) & x > x_0 \end{cases} \quad (1)$$

It yields, indeed, a very good fit of the data (see Fig. 1), *provided that x* is identified with the absolute value of the density, $x = n$. It is easy to show, instead, that Eq. (1) cannot be consistent with experimental data if we choose the variable $x$ equal to $\delta n$.

We introduce, hence, a physically relevant distinction between density $n$ and density fluctuation $\delta n$. This allows us to stress a point, rather trivial in itself, but that deserves to be mentioned: the visualization and investigation of fluctuating quantities in terms of differences from their mean value is a legacy from the Central Limit Theorem (CLT), which states that the PDF for any (macroscopical) $Z$ quantity that can be written as a sum of (microscopical) $z$ stochastic variables, converges to a gaussian, provided that the $z$ variables be independent between them and at least their first two moments be finite. The departure from normality is usually seen as a signature of the existence of correlations in the microscopical dynamics that make the $z$ variables not all independent between them. Within this picture, the mean value $<z>$ is simply an offset of no physical relevance. We want to point out that, in the case of $z$ = density, there is an obvious correlation, not stemming from any dynamics but from the simple requirement for this quantity to be positive definite. This constraint does not play any role as long as one limits to small fluctuations around the mean value, but must radically modify the PDF at least for large negative deviations, which is exactly one of the regions we are investigating now. For a similar caveat about the naïve use of CLT in turbulence, see Ref. [18].

Summarizing, we think that-in the case of the density-it is misleading studying the PDF in terms of just its fluctuations $\delta n$. Instead, also its absolute value is a physically relevant variable. Notice that this reasoning is dictated just by an extremely general principle. As such, they might apply as well to other positive-definite quantities, such as temperature. Indeed, an analysis of PDF $P(T)$ is outside the scope of this work, but some partial investigations have been carried on, and confirm these statements.

From now on, hence, we will not refer any longer to $\delta n$, but only to $n$. Keeping in mind the above paragraphs, and guided by the visual inspection of the empirical PDF-now $P_n(n)$, we are led to argue that a plausible first choice for $P_n$ is the log-normal function:

$$F(n) = \frac{F_0}{n_0} \frac{\exp\left[-\frac{1}{2}\left(\frac{\ln(n/n_0)}{\sigma_n}\right)^2\right]}{n/n_0} \qquad (2)$$

where $F_0$, $n_0$, $\sigma_n$ are free parameters for fitting.

In Fig. 1 we plot the best fit of data using Eq. (2). The $\chi^2$-test yields almost the same goodness of fit either using Eq. (2) or Eq. (1). The agreement with the data is remarkable, even though worsens at the highest values of $n$. This suggests that, perhaps, Eq. (2) is a first-order approximation of an even more refined functional form. We shall discuss this point in the next section.

**IV. An interpretation for the log-normal form for $P_n$**

It is straightforward to notice that a log-normal PDF for a stochastic variable $X$ is the same as a gaussian PDF for the variable $\ln(X)$. It is therefore natural to ask: is it possible to think of some plasma parameter $P$ such that $P = \ln(n)$ and $P$ can be regarded as a stochastic variable with gaussian PDF?

A very rough example is given by the simple zero-dimensional picture of adiabatic electrons, where $n$ is related to the plasma potential $\phi$ through

$$n = n_0 \exp(-\phi/T) \rightarrow \frac{\phi}{T} = -\ln\left(\frac{n}{n_0}\right) \tag{3}$$

Indeed, potential fluctuations in RFX are quite-although not exactly-gaussian[19]. The picture (3) is, however, too simplified when compared with real experimental scenarios (apart for the obvious fact that it deals with electron density and not ion density). We mention just a few references: 1) in RFX, the rms of $\phi/T$ fluctuations, $\Delta_{\phi/T}$, stays almost constant when entering the plasma, while the same quantity for $n$, $\Delta_n$, decreases[20]; 2) in RFX[17] and Frascati Tokamak Upgrade[21] (FTU), measurements of the cross-correlations ($\Delta n_t$, $\Delta \phi_t$) yielded rather low values ( ~ 0.5), not compatible with (Eq. 3); 3) finally, in a review devoted to this subject, Endler[22] remarked that several experiments were confirming that the relative phase angle of the fluctuations in $\phi$ and $n$ lies between $\pi/4$ and $\pi/2$. This, too, is at a variance with (3).

A more refined equation has been developed independently by several groups to relate particles and potential in the SOL. We refer here, e.g., to Krasheninnikov *et al*[23,24] and Sarazin *et al*[25,26]. We report it in the form (Eq. 5 in ref. [24]):

$$n\frac{d}{dt}\nabla_\perp^2 \Phi = \alpha n T^{1/2}\left[1 - e^{-\phi/T}\right] + 2\mathbf{b}\times\mathbf{k}\cdot p \tag{4}$$

where $\Phi \approx 3\,T + \phi$, $p = nT$ is the pressure, $\mathbf{b}$ the unit vector of the magnetic field, $\mathbf{k}$ the magnetic curvature $\mathbf{k} = \mathbf{b}\cdot\nabla\mathbf{b}$, $\alpha = 2\,\rho_s/L$, $\rho_s$ Larmor radius, $L$ the connection length of the field line; $\nabla_\perp$ is the gradient perpendicular to the magnetic field and $d/dt$ is the total time derivative. We also write $2\,\mathbf{b}\times\mathbf{k} = -\beta\,\nabla_y$, with $\beta = 2\,\rho_s/R$ and $R$ the major radius: $\beta$ measures the strength of the curvature drift. The axis $y$ is chosen perpendicular to the magnetic field: it is the poloidal direction in a tokamak and the toroidal one in a RFP. Basically, Eq. (4) is the charge continuity equation for a conducting fluid where an electric field arises because of polarization effects: the driver, here, being the curvature. Numerical coefficients may slightly differ between approaches, depending on the degree of accuracy retained, as well as supplementary terms may enter Eq. (4)-for example, in Refs. [25,26] the viscosity is retained. However, all the essential physics is already there. The lhs is nothing but the time derivative of the net charge density, the first term in the rhs is simply the net current, and the latter term is the polarization drift. Eq. (4), hence, is a minimalist picture for the motion of a conducting fluid in presence of drifts. It can be further simplified into a form suitable to our purposes under the further simplifying assumption that the temperature is almost constant.

It is known from analytical as well as numerical calculations that, among the solutions of Eq. (4), are coherent stable structures (dubbed blobs by Krasheninnikov *et al*) that move rigidly with constant velocity on the top of background plasma. The lhs of Eq. (4) can be cancelled by shifting to the blob's reference frame. Eq. (4) reads, thus,

$$\alpha T^{1/2}\left[1-e^{-\phi/T}\right]-\beta T\frac{\nabla_y n}{n} \equiv \alpha T^{1/2}\left[1-e^{-\phi/T}\right]-\beta T\nabla_y \ln(n/n_0) = 0 \quad (5)$$

where $n_0$ is a reference density used to make dimensionless the argument of the logarithm. We take in Eq. (5) $n$ and $\phi$ as variables, and solve for one of them in terms of the other:

$$\frac{\phi}{T} = -\ln\left[1-\frac{\nabla_y \ln(n/n_0)}{K'}\right] \quad ; \quad \frac{1}{K'} = \frac{\beta}{\alpha}T^{1/2} \quad (6)$$

The presence of the spatial derivative accounts for the phase shift between potential and density fluctuations noticed by Endler[22]. When one looks at absolute values, it is convenient to replace the spatial derivative with an average length (typical size of the blob), $\nabla_y \approx 1/L_y$, and Eq. (6) becomes

$$\frac{\phi}{T} = -\ln\left[1-\frac{\ln(n/n_0)}{K}\right] \quad ; \quad \frac{1}{K} = \frac{\beta}{\alpha}T^{1/2}\frac{1}{L_y} \quad (7)$$

Now, we assume a gaussian PDF for $\phi$, $\hat{F}(\phi)$, and use the conservation of probability through the transformations of variables:

$$\hat{F}(\phi)d\phi = \hat{F}(\phi(n))\frac{d\phi(n)}{dn}dn \equiv \hat{F}(n)\frac{d\phi(n)}{dn}dn$$

$$\rightarrow \hat{F}(n) = \hat{F}_0 \exp\left\{-\frac{1}{2}\frac{\left[\ln\left(1-\frac{\ln(n/n_0)}{K}\right)\right]^2}{\hat{\sigma}_n^2}\right\}\frac{1}{1-\frac{\ln(n/n_0)}{K}}\frac{1}{(n/n_0)} \quad (8)$$

The parameter $\hat{\sigma}_n$ comes from the width of the gaussian for $\phi$. Expression (8) appears rather different from the lognormal curve we guessed at the start. Indeed, the two curves coincide in the limit of small $\varepsilon = \ln(n/n_0)/K$: $\ln(1-\varepsilon) \approx -\varepsilon$ and $1/(1-\varepsilon) \approx 1$. It is formally equivalent to expanding Eq. (6) in powers of $\phi/T$ for $\phi/T \ll 1$, and retaining up to first order (Notice, however, that in RFX potential fluctuations are not small), with the result

$$\ln\left(\frac{n}{n_0}\right) \approx K\frac{\phi}{T} \tag{9}$$

This yields the sought functional relationship between $\ln(n)$ and $\phi$ which, together with the usual ansatz of a normal PDF for $\phi$, yields the log-normal expression for $n$. Expressions (8) or (9) provide the needed decoupling between the radial profiles of the rms of density and potential fluctuations. Infact, even retaining constant in space $L_y$ and $\alpha/\beta = R/L$, there still remains a $T^{-1/2}$ dependence in the $K$ parameter. In deriving Eq. (5) we supposed $T$ spatially constant; however, we may relax this constraint and allow it to be only very weakly varying over the blob's typical size. Since $T$ increases while going towards the centre, it provides the decrease of density fluctuations (notice that, in this case, we are talking about fluctuations). As an example, we provide in Fig. 3 the radial profile of $\Delta_n$. By assuming constant $\Delta_{\phi/T}$ along the radius, as experimentally found[20], Eq. (9) yields $\Delta_n \propto T^{-1/2}$. Notice that it is $\Delta_{\ln(n)}$, not $\Delta_n$, that is proportional to $T^{-1/2}$; but for small density fluctuations, the difference is negligible. By assuming a profile $T(r) = T_{edge} + T_{core}(1-(r/a)^4)$, which is rather typical of RFX, we can try in Fig. 3 a best fit of $\Delta_n$ data.

In Fig. 1 we have provided both fits, using the full PDF (Eq. 8) and the log-normal PDF arising from the truncated expression (Eq. 9). The differences appear fairly limited, although a small improvement appears detectable using the full expression. The linearization of Eq. (5) seems therefore already to retain all the physics needed. However, this is not always the case. In Fig. 4 we plot the same kind of data as of Fig. 1 but for two other RFX shots. In both cases, and unlike Fig. 1, the probe is deeply inserted (about 10 mm). One could wonder if Eq. (4) still holds for such a deep insertion, or rather one is leaving the SOL. We think that the first hypothesys is still valid: in the Reversed Field Pinch there is not a clear distinction between confinement plasma and SOL because of the lack of the separatrix. The presence of large helical deformations of the plasma column in all of RFX discharges makes the rôle of the parallel connection length still important even at such insertion. Also, in Fig. 4(a) we are featuring a pulse with Helium as working gas, while in 4(b) it is standard Hydrogen. Hence, by comparing simultaneously Figs. 1, 4(a), 4(b), we are investigating the potential effects of *two* parameters over the shape of $P_n(n)$: the former parameter is the distance from the wall; the second, the working gas. Notice that the three discharges in Figs. 1, 4(a), 4(b) do feature also some other differences in mean plasma parameters: the mean density is not the same for all (mean temperature, instead, does vary only very slightly). However, we do not think that this can make a large difference. *A priori,* this can be inferred from the fact that the shape of the PDF is universal, regardless of the device, and that different devices feature widely different plasma conditions. The choice of the working gas is, instead, not so trivial since, to the best of our knowledge, all published data refer to experiments done using hydrogen isotopes.

Let us therefore switch back to Fig. 4(a). The empirical PDF is fitted using Eq. (8): the lognormal approssimation (not shown) could not fit the data. Using Eq. (8), instead, allows a roughly satisfactory interpolation, although the quality of the fit is not excellent. In particular, above approximately *n* = 4 (in the units used in the plot), the slopes of the empirical and analytical PDFs begin to severely diverge.

By contrast in the Hydrogen discharge, with the same probe insertion, Fig. 4(b), both fits are plotted, but are practically undistinguishable and both yield good results. Again, the theoretical and empirical curves begin to diverge for $n \geq 4.5 \div 5$, but much less severely. We cannot point to a definite reason for this discrepancy. It may call eventually for a failure of the model which, however, we have shown is able to catch most of the physics running.

As long as we limit to not too large fluctuations, we can notice a close agreement between shallow (figure 1) and deep insertion (figure 4b). This could have been foreseen: since the same curves have been found on different experiments, featuring different probe insertion and SOL widths, the exact radial position must not be a critical parameter in deciding the form of the PDF: if the log-normal approximation works well for a given value of probe insertion, it is likely to work equally well for another value, provided the difference between the two positions is small with respect to the SOL width.

Putting aside the question of the validity of Eq. (5), what could be the cause of the differences between Figs. 4(a) and 4(b) ? We must state in advance that in the following we are just advancing conjectures, since no enough data are available to choose a definite answer. Of course, a first easy explanation is that some terms neglected in Eq. (5) start appearing at this stage, and their relevance be greater in Helium discharges. However, let us see if, already with the informations at hand, it is possible to guess some reasons for the differences. A first plausible comment is that the differences reside in the typical value of $\ln(n/n_0)/K$ : it must be higher in the former case. Since $n_0$ is just a normalization value for the density, the difference must stay in the coefficient $K$, defined in Eq. (8). The edge temperature is not dramatically varying between Hydrogen and Helium plasmas in RFX. The ratio $\alpha/\beta$ depends from geometrical factors which, however, may be different between two shots: since RFP discharges are not so well controlled as typical Tokamaks one, plasma shift may be different in the two shots, thus affecting the value of the parameter $L$. A further possible explanation is related to a variation of the characteristic length $L_y$: it must be lower in Helium plasmas than Hydrogen ones. In the previous discussion we suggested that an interpretation for $L_y$ may be attempted in terms of typical size of density structures. It is also plausible that smaller structures give a reduced net outward flux. Hence, this would correspond to an improved confinement, within this picture. Indeed, the matter is somewhat controversial: an inspection of RFX database tends to ascribe to Helium discharges better confinements properties than Hydrogen ones: the corresponding points are located slightly better on a Greenwald plot[27]. On the other hand, no clear hint of reduced transport comes from direct edge measurements[28].

**V. Conclusions**

We think to have outlined in this work a plausible explanation for the peculiar shape of density PDF at the edge of magnetic fusion devices. Quite remarkably, the explanation starts from a set of equation written for the Tokamak SOL physics. Although RFP's and Tokamaks' edges share many affinities, several differences also occur. Notwithstanding this, the final result is shown not only to describe equally well results for both devices under standard operating conditions, but also to give account for some scenarios (Helium RFP discharges) that, to our knowledge, never appeared in literature before.

Within the picture here outlined, density plays essentially the role of a passive scalar advected by potential fluctuations. A more comprehensive description should take into account also feedback effects of density on potential within a full set of coupled equations. However, the fact that even this simplified approach is able to match empirical data, leads us to think that these effects are small. Hence, the dynamics of the plasma edge is likely to be governed just by a few control parameters: plasma potential and, of course, magnetic field among them. Any attempt of explaining the dynamics of these control parameters is outside the scope of this work.

**Acknowledgments**

Thanks are due to V. Antoni, L. Marrelli and G. Serianni for reading drafts of this paper and raising some important comments.

**Figures**

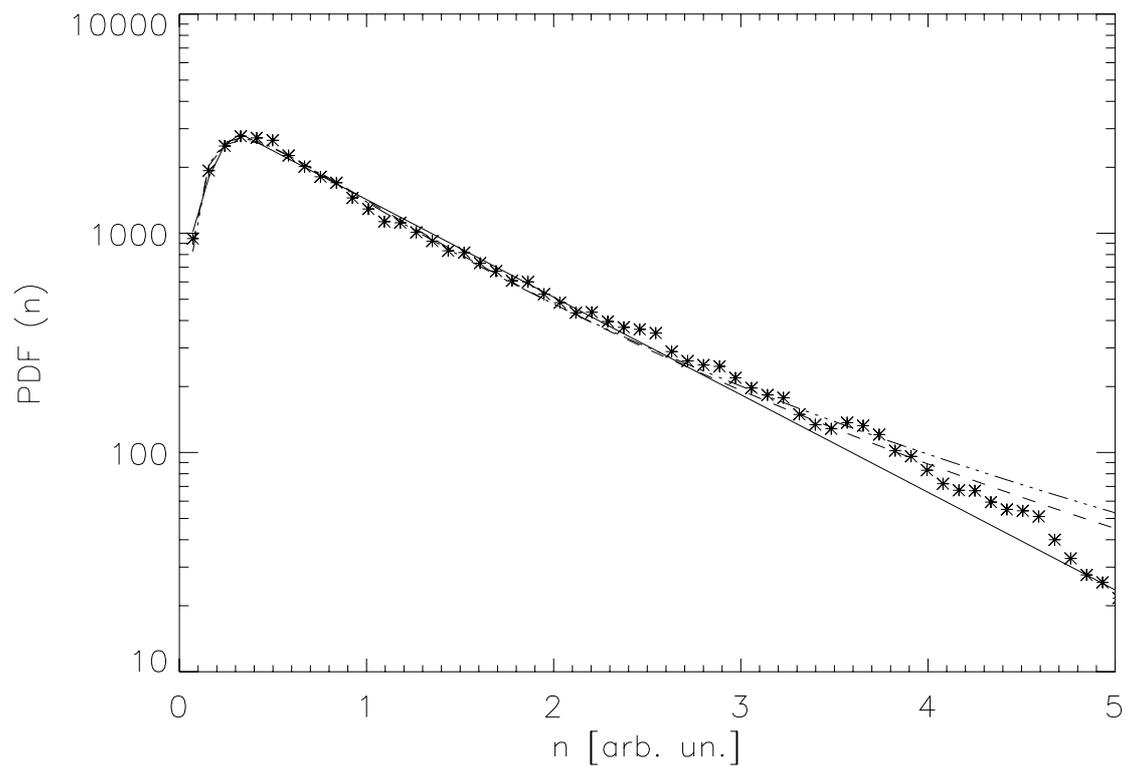

Fig. 1. Unnormalized PDF of density (saturation current) from Langmuir probes. Stars, RFX data points; solid curve, best fit using Eq. (1); chain curve, best fit using Eq. (2); dashed curve, best fit using Eq. (8).

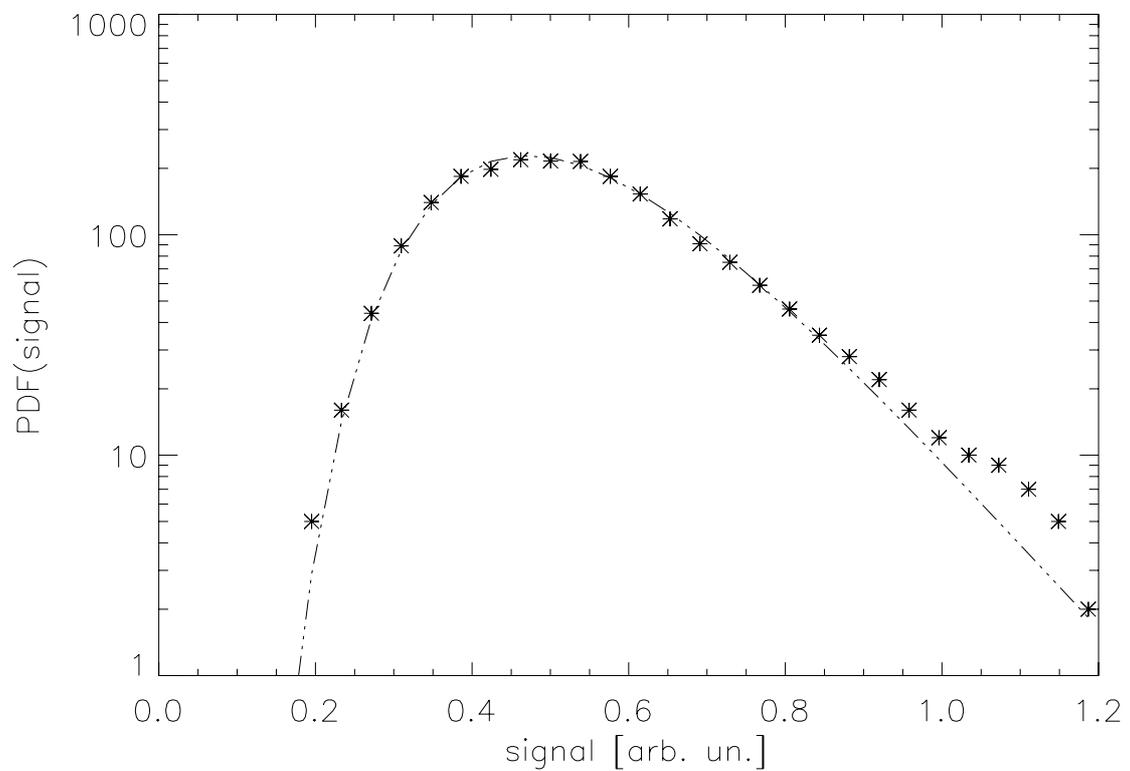

Fig. 2. PDF for CII 5150 Å line signal for a RFX pulse. Stars, data points. Chain curve, best fit using Eq. (2).

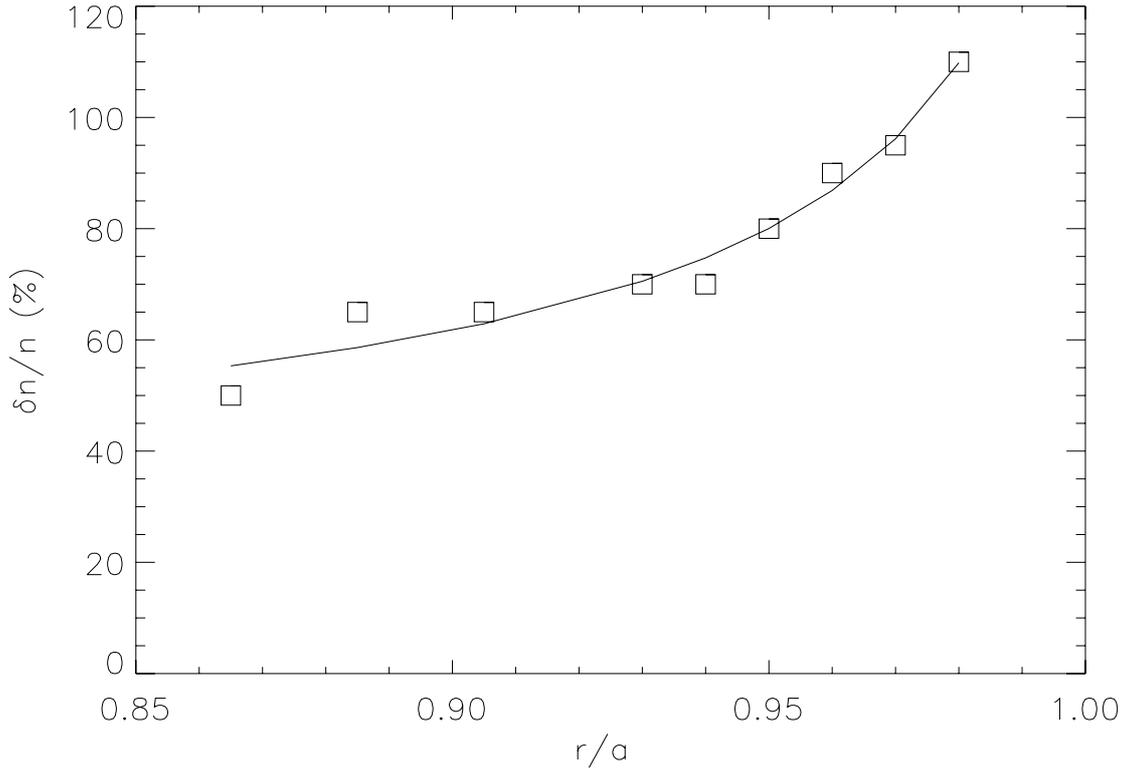

Fig. 3. Normalized mean amplitude of density fluctuations versus radius. Squares are averages over several shots and are adapted from reference [20]. Solid line is a best fit with a curve $C/T^{1/2}$, with $T(r) = T_{edge} + T_{core}(1-(r/a)^4)$. The fit yields $T_{core}/T_{edge} \approx 20$, which is in excellent agreement with RFX profiles.

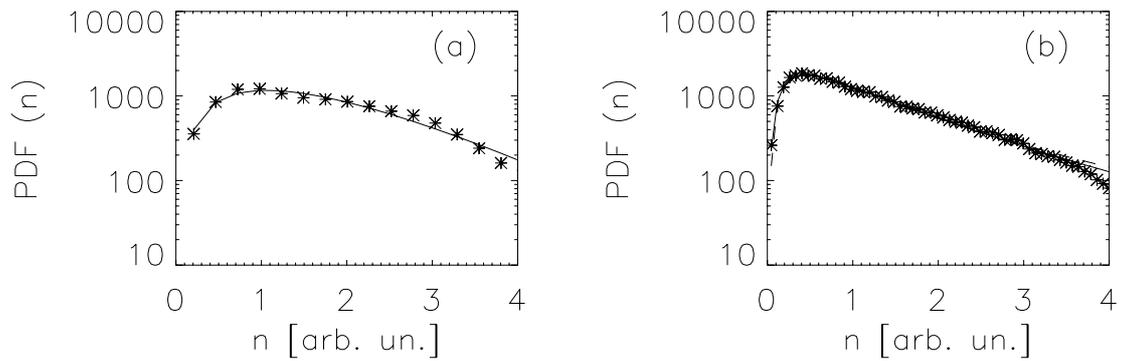

Fig.4. On the left, unnormalized PDF(n) for a Helium RFX pulse with deep (about 1 cm) insertion of the probe. Solid line is the fit from Eq. (8). On the right, the same but for a standard Hydrogen discharge. Here, also the fit using log-normal curve is shown (dashed curve), but the two fits overlap, and are not clearly discernible.